\documentclass[aps,prd,twocolumn,superscriptaddress,showpacs,preprintnumbers,amsmath,amssymb]{revtex4}

\usepackage{graphicx} 
\usepackage{dcolumn}  

\graphicspath{{ps}}

\begin{document}

\title{Measurements of Branching Fractions and Time-dependent {\boldmath$CP$} Violating Asymmetries in {\boldmath$B^{0} \to D^{(*)\pm}D^{\mp}$} Decays}

\affiliation{University of Bonn, Bonn}
\affiliation{Budker Institute of Nuclear Physics SB RAS and Novosibirsk State University, Novosibirsk 630090}
\affiliation{Faculty of Mathematics and Physics, Charles University, Prague}
\affiliation{University of Cincinnati, Cincinnati, Ohio 45221}
\affiliation{Department of Physics, Fu Jen Catholic University, Taipei}
\affiliation{Hanyang University, Seoul}
\affiliation{University of Hawaii, Honolulu, Hawaii 96822}
\affiliation{High Energy Accelerator Research Organization (KEK), Tsukuba}
\affiliation{Hiroshima Institute of Technology, Hiroshima}
\affiliation{Indian Institute of Technology Guwahati, Guwahati}
\affiliation{Indian Institute of Technology Madras, Madras}
\affiliation{Institute of High Energy Physics, Chinese Academy of Sciences, Beijing}
\affiliation{Institute of High Energy Physics, Vienna}
\affiliation{Institute of High Energy Physics, Protvino}
\affiliation{Institute for Theoretical and Experimental Physics, Moscow}
\affiliation{J. Stefan Institute, Ljubljana}
\affiliation{Kanagawa University, Yokohama}
\affiliation{Institut f\"ur Experimentelle Kernphysik, Karlsruher Institut f\"ur Technologie, Karlsruhe}
\affiliation{Korea Institute of Science and Technology Information, Daejeon}
\affiliation{Korea University, Seoul}
\affiliation{Kyungpook National University, Taegu}
\affiliation{\'Ecole Polytechnique F\'ed\'erale de Lausanne (EPFL), Lausanne}
\affiliation{Faculty of Mathematics and Physics, University of Ljubljana, Ljubljana}
\affiliation{Luther College, Decorah, Iowa 52101}
\affiliation{University of Maribor, Maribor}
\affiliation{Max-Planck-Institut f\"ur Physik, M\"unchen}
\affiliation{University of Melbourne, School of Physics, Victoria 3010}
\affiliation{Graduate School of Science, Nagoya University, Nagoya}
\affiliation{Kobayashi-Maskawa Institute, Nagoya University, Nagoya}
\affiliation{Nara Women's University, Nara}
\affiliation{National Central University, Chung-li}
\affiliation{National United University, Miao Li}
\affiliation{Department of Physics, National Taiwan University, Taipei}
\affiliation{H. Niewodniczanski Institute of Nuclear Physics, Krakow}
\affiliation{Nippon Dental University, Niigata}
\affiliation{Niigata University, Niigata}
\affiliation{University of Nova Gorica, Nova Gorica}
\affiliation{Osaka City University, Osaka}
\affiliation{Pacific Northwest National Laboratory, Richland, Washington 99352}
\affiliation{Panjab University, Chandigarh}
\affiliation{Research Center for Nuclear Physics, Osaka University, Osaka}
\affiliation{University of Science and Technology of China, Hefei}
\affiliation{Seoul National University, Seoul}
\affiliation{Sungkyunkwan University, Suwon}
\affiliation{School of Physics, University of Sydney, NSW 2006}
\affiliation{Tata Institute of Fundamental Research, Mumbai}
\affiliation{Excellence Cluster Universe, Technische Universit\"at M\"unchen, Garching}
\affiliation{Toho University, Funabashi}
\affiliation{Tohoku Gakuin University, Tagajo}
\affiliation{Tohoku University, Sendai}
\affiliation{Department of Physics, University of Tokyo, Tokyo}
\affiliation{Tokyo Institute of Technology, Tokyo}
\affiliation{Tokyo Metropolitan University, Tokyo}
\affiliation{Tokyo University of Agriculture and Technology, Tokyo}
\affiliation{CNP, Virginia Polytechnic Institute and State University, Blacksburg, Virginia 24061}
\affiliation{Wayne State University, Detroit, Michigan 48202}
\affiliation{Yamagata University, Yamagata}
\affiliation{Yonsei University, Seoul}
  \author{M.~R\"ohrken}\affiliation{Institut f\"ur Experimentelle Kernphysik, Karlsruher Institut f\"ur Technologie, Karlsruhe} 
  \author{I.~Adachi}\affiliation{High Energy Accelerator Research Organization (KEK), Tsukuba} 
  \author{H.~Aihara}\affiliation{Department of Physics, University of Tokyo, Tokyo} 
  \author{D.~M.~Asner}\affiliation{Pacific Northwest National Laboratory, Richland, Washington 99352} 
  \author{V.~Aulchenko}\affiliation{Budker Institute of Nuclear Physics SB RAS and Novosibirsk State University, Novosibirsk 630090} 
  \author{T.~Aushev}\affiliation{Institute for Theoretical and Experimental Physics, Moscow} 
  \author{A.~M.~Bakich}\affiliation{School of Physics, University of Sydney, NSW 2006} 
  \author{M.~Barrett}\affiliation{University of Hawaii, Honolulu, Hawaii 96822} 
  \author{K.~Belous}\affiliation{Institute of High Energy Physics, Protvino} 
  \author{V.~Bhardwaj}\affiliation{Nara Women's University, Nara} 
% ...
\collaboration{The Belle Collaboration}

\begin{abstract}
We report measurements of branching fractions and time-dependent $CP$ asymmetries in $B^{0} \to D^{+}D^{-}$ and
$B^{0} \to D^{*\pm}D^{\mp}$ decays using a data sample that contains $( 772 \pm 11 )\times 10^6 B\bar{B}$
pairs collected at the $\Upsilon(4S)$ resonance with the Belle detector
at the KEKB asymmetric-energy $e^+ e^-$ collider.
We determine the branching fractions to be
$\mathcal{B}\left(B^{0} \to D^{+}D^{-}\right)= \left( 2.12 \pm 0.16 \pm 0.18 \right)\times 10^{-4}$
and
$\mathcal{B}\left(B^{0} \to D^{*\pm}D^{\mp}\right)= \left( 6.14 \pm 0.29 \pm 0.50 \right)\times 10^{-4}$.
We measure $CP$ asymmetry parameters
$\mathcal{S}_{D^{+}D^{-}} = -1.06_{-0.14}^{+0.21} \pm 0.08$ and
$\mathcal{C}_{D^{+}D^{-}} = -0.43 \pm 0.16 \pm 0.05$
in $B^{0} \to D^{+}D^{-}$
and
$\mathcal{A}_{D^{*}D}       =  +0.06 \pm 0.05 \pm  0.02$,
$\mathcal{S}_{D^{*}D}       =  -0.78 \pm 0.15 \pm  0.05$,
$\mathcal{C}_{D^{*}D}       =  -0.01 \pm 0.11 \pm  0.04$,
$\Delta\mathcal{S}_{D^{*}D} =  -0.13 \pm 0.15 \pm  0.04$ and
$\Delta\mathcal{C}_{D^{*}D} =  +0.12 \pm 0.11 \pm  0.03$
in $B^{0} \to D^{*\pm}D^{\mp}$, where the first uncertainty is statistical and the second is systematic.
We exclude the conservation of $CP$ symmetry in both decays at equal to or greater than $4\sigma$ significance.
\end{abstract}


\maketitle

\tighten

{\renewcommand{\thefootnote}{\fnsymbol{footnote}}}
\setcounter{footnote}{0}
In the standard model (SM) of electroweak interactions, the effect of $CP$ violation is
explained by a single complex phase in the three-family Cabibbo-Kobayashi-Maskawa (CKM)
quark-mixing matrix~\cite{CabibboKobayashiMaskawa}.
Both the Belle and BaBar Collaborations experimentally established this effect~\cite{CPV_observation_Belle,CPV_observation_BaBar}
and precisely determined the parameter $\sin{2\phi_1}$ by measurements of 
mixing-induced $CP$ asymmetries in $b\to(c\bar{c})s$ transitions,
where $\phi_1=\arg\left[-V^{}_{cd} V^{*}_{cb} / V^{}_{td} V^{*}_{tb} \right]$~\cite{BaBar_btoccs,Belle_btoccs, phione}.

In $b\to c\bar{c}d$ transitions such as $B^{0} \to D^{(*)\pm}D^{\mp}$ decays, the dominant contributions are Cabibbo-disfavored
but color-allowed tree-level diagrams and the corresponding mixing-induced $CP$ asymmetries are
directly related to $\sin{2\phi_1}$.
In addition, $b \to d$ penguin diagrams that may have different weak phases can contribute to these decays. 
Theoretical considerations based on models using factorization approximations and heavy quark symmetry
predict the corrections to mixing-induced $CP$ violation to be a few percent and possible direct $CP$ violation
to be negligibly small~\cite{Xing1998Xing1999}.

$CP$ violation in $b\to c\bar{c}d$ transitions has been studied previously by
the Belle and BaBar Collaborations.
In $B^{0} \to D^{+}D^{-}$ decays
using a data sample of $535 \times 10^6 B\bar{B}$ pairs,
Belle found evidence of a large direct $CP$ violation:
$\mathcal{C}_{D^{+}D^{-}} = -0.91 \pm 0.23 \pm 0.06$
corresponding to a $3.2\sigma$ deviation from zero~\cite{Fratina2007,directCPV_A_or_C},
in contradiction to theoretical expectations~\cite{Xing1998Xing1999}.
This deviation was not confirmed by BaBar and has not been
observed in other $B^{0} \to D^{*\pm}D^{(*)\mp}$ decay modes~\cite{Aushev2004, Vervink2009, Aubert2009_btodd}.

In this article we present measurements of branching fractions and $CP$ violating asymmetries in
the decays $B^{0} \to D^{+}D^{-}$ and $B^{0} \to D^{*\pm}D^{\mp}$ using the final data sample of the Belle experiment.

The decay rate of a neutral $B$ meson decaying to a $CP$ eigenstate such as $D^{+}D^{-}$
is given by
\begin{align}
  f_{D^{+}D^{-}} ( \Delta t ) = 
  \frac{ e^{ - \lvert \Delta t \rvert / \tau_{B^{0}} } }{ 4 \tau_{B^{0}} } &
  \lbrace
  1 + q [
  \mathcal{S}_{D^{+}D^{-}} \sin( \Delta m_{d} \Delta t ) \nonumber \\
  & - \mathcal{C}_{D^{+}D^{-}}  \cos( \Delta m_{d} \Delta t )
  ] \rbrace , \label{decay_rate_btodd}
\end{align}
where $q = +1 \,(-1)$ represents the $b$-flavor charge when the accompanying $B$ meson
is tagged as a $B^{0}$ ($\bar{B}^{0}$), and $\Delta t$ represents the
proper time interval between the two neutral $B$ decays in an $\Upsilon\left(4S\right)$ event.
The $B^{0}$ lifetime is denoted by $\tau_{B^{0}}$ and the mass difference between the
two neutral $B$ mass eigenstates by $\Delta m_{d}$. The parameters $\mathcal{S}_{D^{+}D^{-}}$
and $\mathcal{C}_{D^{+}D^{-}}$ measure mixing-induced and direct $CP$ violation, respectively~\cite{directCPV_A_or_C}.

Unlike $D^{+}D^{-}$, $D^{*+}D^{-}$ and $D^{*-}D^{+}$ are not $CP$ eigenstates.
The decay rate of neutral $B$ mesons decaying to these states has
four flavor-charge configurations and can be expressed as~\cite{Aleksan1991,Aubert2003_b_to_rhopi_parametrisation}
\begin{align}
  f_{D^{*\pm}D^{\mp}} ( \Delta t ) = 
  ( 1 \pm \mathcal{A}_{D^{*}D} ) & 
  \frac{ e^{ - \lvert \Delta t \rvert / \tau_{B^{0}} } }{ 8 \tau_{B^{0}} } \nonumber \\
  \times \lbrace
  1 
  + q [ 
  ( \mathcal{S}_{D^{*}D} \pm & \Delta\mathcal{S}_{D^{*}D} ) \sin( \Delta m_{d} \Delta t ) \nonumber \\
  - ( \mathcal{C}_{D^{*}D} \pm & \Delta\mathcal{C}_{D^{*}D} ) \cos( \Delta m_{d} \Delta t )
  ] \rbrace , \label{decay_rate_btodstard}
\end{align}
where the $+$ ($-$) sign represents the $D^{*+}D^{-}$ ($D^{*-}D^{+}$) final state.
The time- and flavor-integrated charge asymmetry $\mathcal{A}_{D^{*}D}$ measures
direct $CP$ violation.
The quantity $\mathcal{S}_{D^{*}D}$ parameterizes mixing-induced $CP$ violation
and $\mathcal{C}_{D^{*}D}$ parameterizes flavor-dependent direct $CP$ violation.
The quantities $\Delta\mathcal{C}_{D^{*}D}$ and $\Delta\mathcal{S}_{D^{*}D}$ are not sensitive to
$CP$ violation.
The parameter $\Delta\mathcal{C}_{D^{*}D}$ describes the asymmetry between the rates
$\Gamma ( B^{0} \to D^{*-} D^{+}  )  +  \Gamma ( \bar{B}^{0} \to D^{*+} D^{-} )$ and
$\Gamma ( B^{0} \to D^{*+} D^{-} )  +  \Gamma ( \bar{B}^{0} \to D^{*-} D^{+}  )$.
The parameter $\Delta\mathcal{S}_{D^{*}D}$ is related to the relative strong phase between the amplitudes
contributing to the decays.

This analysis is based on a data sample containing $( 772 \pm 11 )\times 10^6 B\bar{B}$
pairs collected at the $\Upsilon(4S)$ resonance with the Belle detector
at the KEKB asymmetric-energy $e^+ e^-$ collider~\cite{KEKB}.
The $\Upsilon(4S)$ is produced with a Lorentz boost of $\beta\gamma = 0.425$ close to an
axis along the $e^{-}$ beam, which allows the determination of $\Delta t$ from the
displacement of decay vertices of both $B$ mesons.

The Belle detector is a large-solid-angle magnetic
spectrometer that is described in detail in Ref.~\cite{Belle}.
The present analysis uses for track reconstruction and particle identification
a silicon vertex detector (SVD),
a 50-layer central drift chamber (CDC), an array of
aerogel threshold Cherenkov counters (ACC),  
a barrel-like arrangement of time-of-flight
scintillation counters (TOF), and an electromagnetic calorimeter
composed of CsI(Tl) crystals (ECL) located inside 
a superconducting solenoid coil that provides a 1.5~T
magnetic field.

Reconstructed charged tracks are required to have a transverse (longitudinal)
distance of closest approach to the interaction point (IP) of less than $2$ ($4$) cm.
For identification of charged particles (PID), measurements of specific energy loss in the CDC
and measurements from the ACC and TOF are combined in an likelihood-ratio approach.
The selection requirement on the combined PID quantity has a kaon (pion) identification efficiency
of $91\%$ ($99\%$) with an associated pion (kaon) misidentification rate of $2\%$ ($18\%$).
Charged tracks are also required to be not positively identified as electrons
by measurements of shower shapes and energy deposited in the ECL.
Neutral pions are reconstructed from two photons detected in the ECL with each photon having
an energy greater than $30~\textrm{MeV}$. The invariant mass of the photon pair is required to
be within $15~\textrm{MeV/}c^2$ of the nominal $\pi^{0}$ mass (corresponding to a width of $3.3\sigma$). For $\pi^{0}$ candidates
a kinematic fit to the IP profile with a mass constraint is performed.
Neutral kaons are reconstructed in the decay mode $K_{S}^{0} \to \pi^{+}\pi^{-}$. The
invariant mass of the $\pi^{+}\pi^{-}$ pair is required to be within $15~\textrm{MeV/}c^2$
of the nominal $K_{S}^{0}$ mass ($5.8\sigma$).
Additional momentum-dependent selection requirements consider the possible displacement
of the $K_{S}^{0}$ decay vertices from the IP~\cite{Kshortselection}.

Charged $D$ mesons are reconstructed in the decay modes $D^{+} \to K^{-} \pi^{+} \pi^{+}$
and $D^{+} \to K_{S}^{0} \pi^{+}$~\cite{CC}.
The invariant mass of $D^{+}$ candidates is required to be
within $12~\textrm{MeV/}c^2$ of the nominal mass ($3.4\sigma$ in $D^{+} \to K^{-} \pi^{+} \pi^{+}$ and $2.9\sigma$ in $D^{+} \to K_{S}^{0} \pi^{+}$).
Neutral $D$ mesons are reconstructed in the decay modes $D^{0} \to K^{-} \pi^{+}$,
$D^{0} \to K^{-} \pi^{+} \pi^{+} \pi^{-}$, $K_{S}^{0} \pi^{+} \pi^{-}$
and $D^{0} \to K^{-} \pi^{+} \pi^{0}$.
The invariant mass of $D^{0}$ candidates is required to be
within $15~\textrm{MeV/}c^2$ ($3.3\sigma-3.7\sigma$) of the nominal mass, except for the
$D^{0} \to K^{-} \pi^{+} \pi^{0}$ decay mode where a 
requirement of $32~\textrm{MeV/}c^2$ ($3.0\sigma$) is applied.
We reconstruct $D^{*+}$ mesons in the decay modes $D^{*+} \to D^{0} \pi^{+}$
and $D^{*+} \to D^{+} \pi^{0}$.
The momentum resolution of charged low momentum pions
from $D^{*+}$ decays, referred to as soft pions, is improved by a kinematic fit in
which the soft pion is constrained to the $D^{*+}$ decay vertex determined 
from a kinematic fit of $D$ candidates constrained to originate
from the IP profile.
The difference of invariant masses of the $D^{*+}$ and $D^{0}$ ($D^{+}$) candidates is required to
be within $1.5~\textrm{MeV/}c^2$ ($2.5~\textrm{MeV/}c^2$) ($3.1\sigma-3.7\sigma$) of the nominal mass difference, 
except for modes involving
$D^{0} \to K^{-} \pi^{+} \pi^{+} \pi^{-}$ and $D^{0} \to K^{-} \pi^{+} \pi^{0}$ decays
where a requirement of $2~\textrm{MeV/}c^2$ ($4.6\sigma$ and $3.2\sigma$) is applied.

Neutral B mesons are reconstructed by combining $D^{(*)+}$ and $D^{-}$ candidates,
and selected by
the beam-energy-constrained mass $M_{\rm bc} = \sqrt{ ( E^{*}_{\rm beam} / c^{2} )^{2} - ( p^{*}_{B} / c )^{2} }$ and
the energy difference $\Delta E = E^{*}_{B} - E^{*}_{\rm beam}$,
where $E^{*}_{\rm beam}$ is the energy of the beam
and $p^{*}_{B}$ and $E^{*}_{B}$ are the momentum and energy of the $B^{0}$ candidates
in the center-of-mass frame (c.m.). The selected regions are 
$5.2~\mathrm{GeV}/\mathrm{c}^{2}<M_{\rm bc}<5.3~\mathrm{GeV}/\mathrm{c}^{2}$
and $-50~\mathrm{MeV}<\Delta E<100~\mathrm{MeV}$. The lower boundary in $\Delta E$ was chosen
to exclude reflections from misidentified $B^{0} \to D_{s}^{+} D^{(*)-}$ decays
that populate the $M_{\rm bc}$ signal region at $\Delta E\approx-75~\mathrm{MeV}$.

In $B^{0} \to D^{+} D^{-}$ ($B^{0} \to D^{*\pm} D^{\mp}$),
after applying the above selection requirements, 12\% (16\%) of the signal events
contain more than one $B^{0}$ candidate.
In this case the candidate with the smallest quadratic sum of deviations
of reconstructed invariant masses of $D$ daughters (and mass differences of $D^{*+}$ daughters)
from nominal values, divided by the width of corresponding signal peaks, is selected.
This requirement selects the correct candidate with a probability of 96\% (92\%).

In $B^{0} \to D^{+} D^{-}$ unlike in $B^{0} \to D^{*\pm} D^{\mp}$ the major source of
background arises from $e^{+}e^{-} \to q \bar{q}$  $( q \in \{ u, d, s, c \} )$ continuum events.
This background is suppressed by a neural network (NN)
implemented by the NeuroBayes package~\cite{Neurobayes}
that combines information about the event topology.
Observables included in the NN are $\cos \theta^{*}_{B}$, where $\theta^{*}_{B}$
is the polar angle of the $B^{0}$ candidate with respect to the beam direction in the c.m. frame,
a combination of 16 modified Fox-Wolfram moments~\cite{FWmoments},
and the momentum flow in nine concentric cones around the thrust axis of the $B^{0}$ candidate~\cite{CLEOcones}.
The requirement on the NN selection rejects 64\% of the background while retaining 92\% of the signal.

\begin{figure}[htb]
\includegraphics[width=0.475\textwidth]{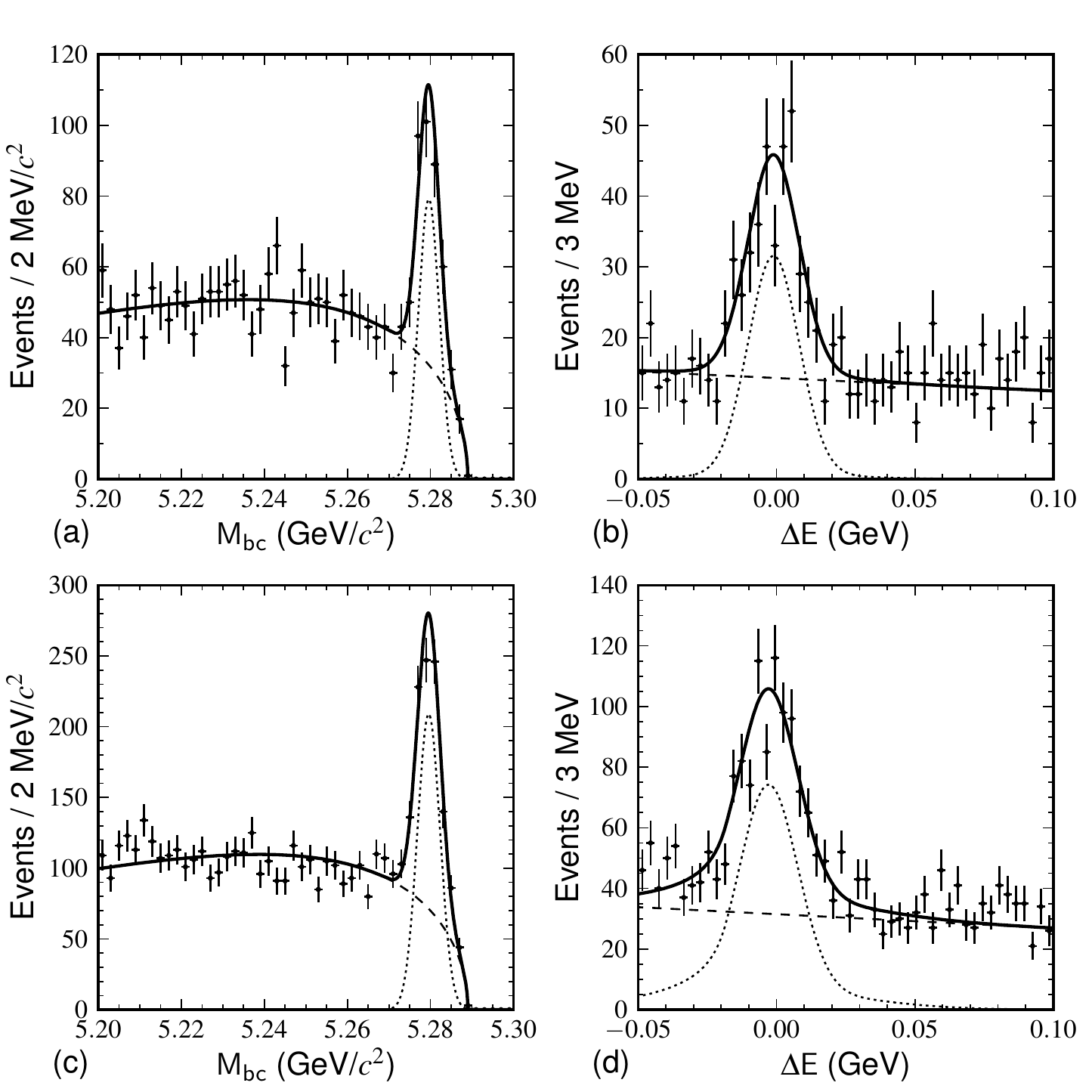}
\caption{
$M_{\rm bc}$ and $\Delta E$ distributions (data points with error bars) and fit projections (solid lines)
for (a)-(b) $B^{0} \to D^{+} D^{-}$ and (c)-(d) $B^{0} \to D^{*\pm} D^{\mp}$ decays .
The dotted (dashed) lines represent projections of signal (background) fit components.
A $\lvert\Delta E\rvert<30~\textrm{MeV}$ ($M_{\rm bc}>5.27~\textrm{GeV/}c^{2}$) requirement is applied in plotting the $M_{\rm bc}$ ($\Delta E$) distributions.
}
\label{figure_mbc_deltae_distributions}
\end{figure}

The signal yields are obtained by two-dimensional unbinned extended maximum likelihood fits
to the $M_{\rm bc}$ and $\Delta E$ distributions.
The $M_{\rm bc}$ distributions are parameterized by a Gaussian function for the
signal component and by an empirically determined threshold function introduced by the
ARGUS Collaboration~\cite{ARGUSfunction} for the background component.
The $\Delta E$ distributions are parameterized by the sum of two Gaussian functions
(the sum of a Gaussian function and an empirically determined function
introduced by the Crystal Ball Collaboration~\cite{CBfunction}) with common mean for the signal component
in $B^{0} \to D^{+} D^{-}$ ($B^{0} \to D^{*\pm} D^{\mp}$)
and by a linear function for the background component.
The shape parameters of signal components in $B^{0} \to D^{+} D^{-}$ ($B^{0} \to D^{*\pm} D^{\mp}$)
are fixed to values obtained from $B^{0} \to D_{s}^{+} D^{-}$ ($B^{0} \to D_{s}^{+} D^{*-}$)
data distributions, where the relative widths and fractions of the signal components in $\Delta E$ are
fixed to values obtained from Monte Carlo (MC) simulation studies.
The $M_{\rm bc}$ and $\Delta E$ distributions and fit projections are shown in Fig.~\ref{figure_mbc_deltae_distributions}.
For $B^{0} \to D^{+} D^{-}$ the obtained yields are $221.4 \pm 18.6$ signal events in the
$(K^{-} \pi^{+} \pi^{+}) (K^{+} \pi^{-} \pi^{-})$ final state and
$48.0 \pm 8.9$ signal events in the $(K^{-} \pi^{+} \pi^{+}) (K_{S}^{0} \pi^{-})$ final state.

For $B^{0} \to D^{*\pm} D^{\mp}$, we obtain a yield of $886.8 \pm 39.3$ signal events
in all reconstructed modes combined. Of these, the yield in modes involving
$D^{*+} \to D^{0} \pi^{+}$ decays only is $769.2 \pm 36.0$ signal events.

Decays such as $B^{0} \to D^{(*)-} K^{*+}$, $B^{0} \to D^{(*)-} K^{0}\pi^{+}$ and $B^{0} \to D^{(*)-} \pi^{+}\pi^{+}\pi^{-}$ have
the same final states as the reconstructed $B^{0} \to D^{+} D^{(*)-}$ decay modes
and can possibly populate the $M_{\rm bc}$ and $\Delta E$ signal region. The contributions of
such decays, referred to as peaking background, are estimated from $D$ mass sidebands
and subtracted in the signal yields given above.
For $B^{0} \to D^{+} D^{-}$ ($B^{0} \to D^{*\pm} D^{\mp}$), we find a contribution of $0.7 \pm 1.5$ ($4.7 \pm 2.1$)
peaking background events from fits to $D^{-} \to K_{S}^{0} \pi^{-}$ mass sidebands.
The $D^{-} \to K^{+} \pi^{-} \pi^{-}$ mass sidebands are considered to be free of peaking background
and no background subtraction is performed. This assumption has been tested by MC simulations
and no peaking background is found in the data sidebands.

The reconstruction efficiencies are obtained from MC simulations of signal decays and
have been corrected to account for PID selection efficiency differences between MC simulations and data.
To exclude systematic effects in the determination of reconstruction efficiencies associated 
with soft neutral pions, only modes involving $D^{*+} \to D^{0} \pi^{+}$ decays are used
in the $B^{0} \to D^{*\pm}D^{\mp}$ branching fraction measurement.

The branching fractions are calculated from signal yields, reconstruction efficiencies,
the number of $B\bar{B}$ events and
current world averages of $D^{0}$, $D^{+}$ and $D^{*+}$ branching fractions~\cite{PDG}.
The branching fraction for $B^{0} \to D^{+}D^{-}$ decays is calculated as the weighted average of
the branching fractions determined for each of both reconstructed decay modes separately.
The branching fraction for $B^{0} \to D^{*\pm}D^{\mp}$ decays is determined by the signal yield
in all modes and the average reconstruction efficiency weighted by the $D$ branching fractions.
The determined branching fractions are
$\mathcal{B}\left(B^{0} \to D^{+}D^{-}\right)= \left( 2.12 \pm 0.16 \pm 0.18 \right)\times 10^{-4}$
and
$\mathcal{B}\left(B^{0} \to D^{*\pm}D^{\mp}\right)= \left( 6.14 \pm 0.29 \pm 0.50 \right)\times 10^{-4}$.
\begin{table}[htb]
\caption{ Summary of systematic uncertainties of the
$B^{0} \to D^{+}D^{-}$ and $B^{0} \to D^{*\pm} D^{\mp}$ branching fractions (in \%).}
\label{systemtic_uncertainties_branching_fraction}
\begin{tabular}{lcc}
\hline \hline
Source & 
$D^{+}D^{-}$ &
$D^{*\pm}D^{\mp}$ \\
\hline
Track reconstruction efficiency                       &   $2.0$    &   $4.1$    \\
$K_{S}^{0}$ reconstruction efficiency                 &   $0.7$    &   $0.7$    \\
$\pi^{0}$ reconstruction efficiency                   &     -      &   $1.6$    \\
$K/\pi$ selection efficiency                          &   $5.5$    &   $5.3$    \\
Event reconstruction efficiency                       &   $1.0$    &   $0.1$    \\
Continuum suppression                                 &   $4.1$    &     -      \\
Fit models                                            &   $1.1$    &   $0.6$    \\
$D$ branching fractions                               &   $4.3$    &   $3.9$    \\
Number of $B\bar{B}$ events                           &   $1.4$    &   $1.4$    \\
\hline
Total                                                 &   $8.6$    &   $8.1$    \\
\hline \hline
\end{tabular}
\end{table}
The systematic uncertainties of the measured branching fractions are summarized
in Table~\ref{systemtic_uncertainties_branching_fraction}.
The uncertainties due to track, $K_{S}^{0}$ and $\pi^{0}$ reconstruction efficiency
and the uncertainty due to the $\mathrm{K}/\pi$ selection efficiency
have been estimated using studies of $D^{*+}$ decays with MC simulations and data.
The effect on the event reconstruction efficiencies due to broader $D$ mass distributions for data
and the corresponding selection is studied by a MC/data comparison and assigned as
a systematic uncertainty.
As the systematic uncertainty of the applied continuum suppression in $B^{0} \to D^{+}D^{-}$, the
maximum variation of signal yields in a MC/data comparison of the neural networks using
$B^{0} \to D_{s}^{+} D^{-}$ decays is assigned.
The contributions due to the fit models are estimated by varying the fixed parameters within their uncertainties.
The contributions due to uncertainties of the $D^{0}$, $D^{+}$ and $D^{*+}$ branching fractions
and of the number of $B\bar{B}$ events are obtained by propagation of the appropriate uncertainties.
The total systematic uncertainties are obtained by adding all contributions in quadrature.

\begin{figure}[htb]
\includegraphics[width=0.475\textwidth]{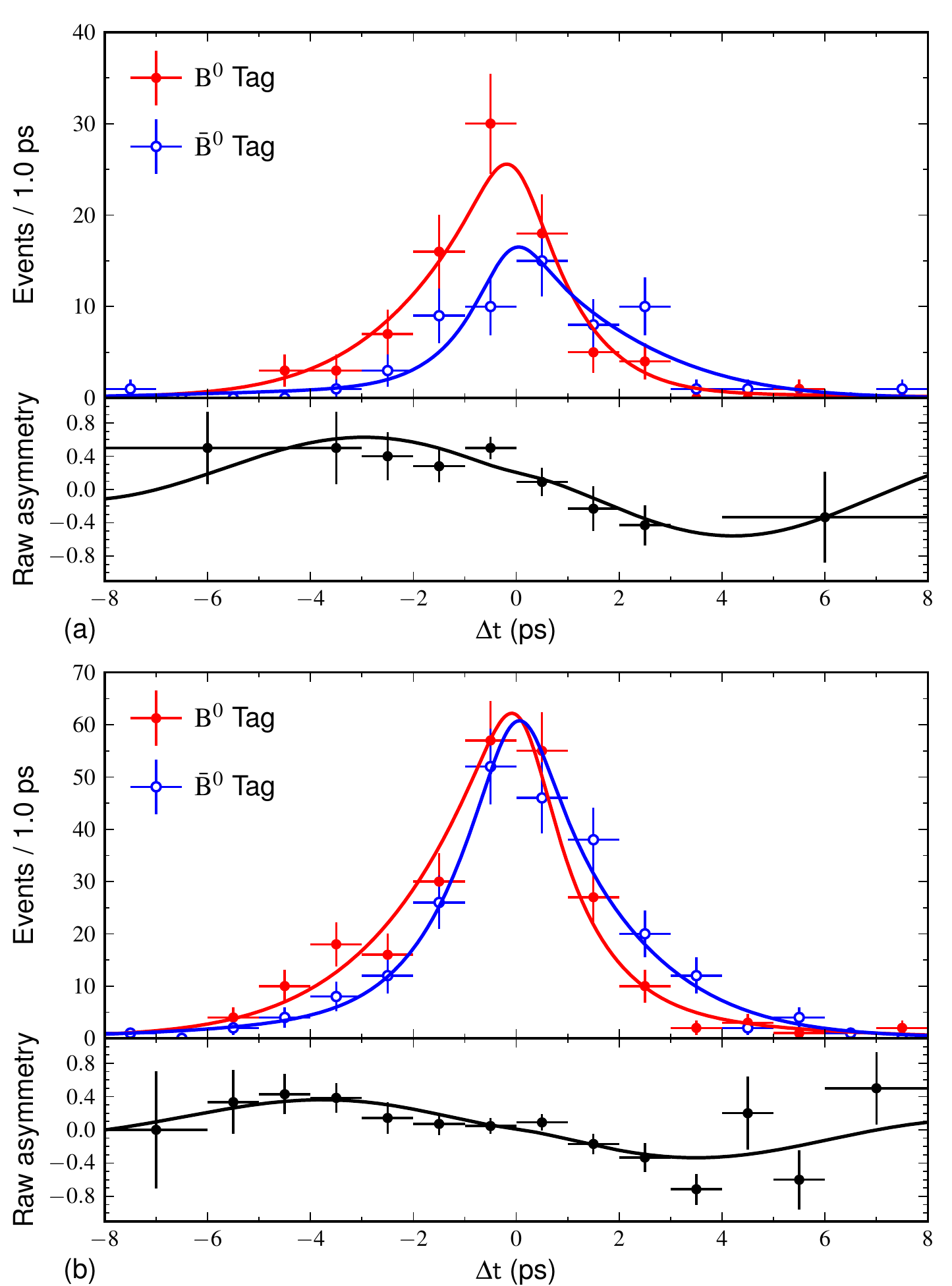}
\caption{
Top: $\Delta t$ distributions (data points with error bars) of (a) $B^{0} \to D^{+} D^{-}$ and (b) $B^{0} \to D^{*+} D^{-} + B^{0} \to D^{*-} D^{+}$ candidates
     associated with high quality flavor tags ($r > 0.5$). The lines show projections of the sum of signal and background components in the fit.
     The signal purity for $r > 0.5$ is 69\% (66\%) for $B^{0} \to D^{+} D^{-}$ ($B^{0} \to D^{*\pm} D^{\mp}$).
Bottom: The $CP$ asymmetry obtained from the above distributions and projections.
}
\label{figure_cpfit}
\end{figure}
\begin{table*}[htb]
\caption{ Summary of systematic uncertainties in the time-dependent $CP$ asymmetry parameters for
$B^{0} \to D^{+}D^{-}$ and $B^{0} \to D^{*\pm} D^{\mp}$ decays (in units of $10^{-2}$).}
\label{systemtic_uncertainties_cpfit}
\begin{tabular}{lccccccc}
\hline \hline
Source & 
$\mathcal{S}_{D^{+}D^{-}}$ &
$\mathcal{C}_{D^{+}D^{-}}$ & 
$\mathcal{A}_{D^{*}D}$       &
$\mathcal{S}_{D^{*}D}$       &
$\mathcal{C}_{D^{*}D}$       &
$\Delta\mathcal{S}_{D^{*}D}$ &
$\Delta\mathcal{C}_{D^{*}D}$ \\
\hline
Vertex reconstruction                                 &   $3.6$    &   $2.2$    &   $1.3$   &   $2.5$   &   $2.3$   &   $2.4$   &   $2.3$  \\  
$\Delta t$ resolution function                        &   $6.5$    &   $2.4$    &   $0.4$   &   $3.5$   &   $1.1$   &   $1.9$   &   $0.6$  \\
Background $\Delta t$ PDFs                            &   $2.7$    &   $0.5$    &   $0.2$   &   $0.7$   &   $0.2$   &   $0.5$   &   $0.1$  \\
Signal purity                                         &   $1.2$    &   $1.8$    &   $0.2$   &   $0.9$   &   $0.4$   &   $0.3$   &   $0.2$  \\
Physics parameters                                    &   $0.7$    &   $0.4$    &  $<0.1$   &   $0.2$   &   $0.1$   &   $0.2$   &  $<0.1$  \\
Flavor tagging                                       &   $0.7$    &   $0.6$    &  $<0.1$   &   $0.4$   &   $0.3$   &   $0.3$   &   $0.2$  \\
Possible fit bias                                     &   $0.8$    &   $0.2$    &   $0.6$   &   $0.8$   &   $1.1$   &   $0.8$   &   $0.5$  \\
Peaking background                                    &   $0.3$    &   $0.9$    &   $0.4$   &   $1.3$   &   $0.5$   &   $0.8$   &   $0.7$  \\
Tag-side interference                                 &   $1.4$    &   $3.2$    &   $0.2$   &   $1.1$   &   $3.1$   &   $0.9$   &   $0.6$  \\
\hline
Total                                                 &   $8.2$    &   $5.1$    &   $1.6$   &   $4.9$   &   $4.3$   &   $3.5$   &   $2.6$  \\
\hline \hline
\end{tabular}
\end{table*}
The technique used to determine the $CP$ asymmetry parameters from $\Delta t$ distributions is
described in detail in Ref.~\cite{Belle_btoccs}.
The decay vertex of the signal $B$ meson is reconstructed from a kinematic fit
of the two $D$ mesons to a common vertex including information about the IP profile.
No information about soft pions is used in the vertex reconstruction.
The decay vertex and the flavor of the accompanying $B$ meson is obtained by an inclusive approach using the
remaining charged tracks that are not used in the signal $B$ reconstruction.
Requirements on the quality of reconstructed $B$ vertices and on the number of hits in the silicon vertex detector
are applied.
The algorithms applied to obtain the $b$-flavor charge $q$ and a tagging quality variable $r$ are described
in detail in Ref.~\cite{TaggingNIM}.
The variable $r$ is related to the mistag fractions determined from $b\to c$ control samples and
ranges from $r=0$ (no flavor discrimination) to $r=1$ (unambiguous flavor assignment).
The data is divided into seven $r$ intervals.

The $CP$ asymmetry parameters are determined by unbinned maximum likelihood fits to the $\Delta t$ distributions.
The probability density function used to describe the $\Delta t$ distributions is given by
\begin{align}
P =
 ( 1 & - f_{\rm ol} )
\sum\limits_{k}
f_{k} 
\int
\left[
\mathcal{P}_{k} \left( \Delta t' \right)
R_{k} \left( \Delta t - \Delta t' \right) 
\right]
d\left( \Delta t' \right ) \nonumber \\
+ & f_{\rm ol} P_{\rm ol} \left( \Delta t \right) ,
\end{align}
where the index $k$ denotes signal and background components and the fraction $f_{k}$ depends on the $r$ interval and
is evaluated on an event-by-event basis as a function of $M_{\rm bc}$ and $\Delta E$.
The signal component consists of the convolution of distributions given
by modifications of Eq.~\ref{decay_rate_btodd} and~\ref{decay_rate_btodstard} that include the effect of
incorrect flavor assignments and of a resolution function to account for the finite resolution of the vertex reconstruction~\cite{vertexres}.
The background component is parameterized by the convolution of the sum of a prompt and an exponential distribution
allowing for effective lifetimes and a resolution function composed of the sum of two Gaussian functions.
The parameters of the background components are fixed to values determined by fits
to $M_{\rm bc}<5.26~\textrm{GeV/}c^{2}$ sidebands.
A Gaussian function $P_{\rm ol}$ with a broad width of about $35$ ps and a small fraction $f_{\rm ol}$ of about $2 \times 10^{-4}$
is added to account for outlier events with large $\Delta t$.

The free parameters in the $B^{0} \to D^{+} D^{-}$ fit are
$\mathcal{S}_{D^{+}D^{-}}$ and $\mathcal{C}_{D^{+}D^{-}}$
and the free parameters in the $B^{0} \to D^{*\pm} D^{\mp}$ fit are
$\mathcal{A}_{D^{*}D}$, $\mathcal{S}_{D^{*}D}$, $\mathcal{C}_{D^{*}D}$, $\Delta\mathcal{S}_{D^{*}D}$ and $\Delta\mathcal{C}_{D^{*}D}$.
The lifetime $\tau_{B^{0}}$ and mass difference $\Delta m_{d}$ are fixed to current world averages~\cite{PDG}.
The fits are performed in a signal region defined by
$\lvert\Delta E\rvert<30~\textrm{MeV}$ and $5.27~\textrm{GeV/}c^{2} < M_{\rm bc} <5.29~\textrm{GeV/}c^{2}$.
The signal purity is 62\% (59\%) for $B^{0} \to D^{+}D^{-}$ ($B^{0} \to D^{*\pm}D^{\mp}$).
For $B^{0} \to D^{+}D^{-}$ the results are
\begin{alignat}{2}
\mathcal{S}_{D^{+}D^{-}} &= -1.06_{\ -0.14}^{\ +0.21} &&\pm 0.08 \nonumber \\
\mathcal{C}_{D^{+}D^{-}} &= -0.43 \pm 0.16 &&\pm 0.05,
\end{alignat}
and for $B^{0} \to D^{*\pm}D^{\mp}$
\begin{alignat}{2}
\mathcal{A}_{D^{*}D}       &=  +0.06 \pm 0.05 &&\pm  0.02 \nonumber \\
\mathcal{S}_{D^{*}D}       &=  -0.78 \pm 0.15 &&\pm  0.05 \nonumber \\
\mathcal{C}_{D^{*}D}       &=  -0.01 \pm 0.11 &&\pm  0.04 \nonumber \\
\Delta\mathcal{S}_{D^{*}D} &=  -0.13 \pm 0.15 &&\pm  0.04 \nonumber \\
\Delta\mathcal{C}_{D^{*}D} &=  +0.12 \pm 0.11 &&\pm  0.03,
\end{alignat}
where the first uncertainty is statistical and the second systematic.
The $\Delta t$ distributions and projections of the fits are shown in Fig.~\ref{figure_cpfit}.

The systematic uncertainties in the $CP$ asymmetry parameters are evaluated for each decay mode
and are summarized in Table~\ref{systemtic_uncertainties_cpfit}. Sources of systematic uncertainties on the vertex reconstruction are
the IP profile constraint, requirements on the vertex fit quality for signal and tagging $B$ mesons,
requirements on impact parameters of tracks in the reconstruction of the tagging $B$ meson and the $\Delta t$ fit range.
These contributions are estimated by variations of each of the applied requirements.
Further contributions to the vertex reconstruction are a global SVD misalignment
and a $\Delta z$ bias, which are both estimated by MC simulations.
The contributions due to
the $\Delta t$ resolution functions,
the $\Delta t$ parameterization of background components,
the calculation of the signal purity
and the physics parameters $\tau_{B^{0}}$ and $\Delta m_{d}$
are estimated by varying the fixed parameters within their uncertainties.
The systematic uncertainty due to flavor tagging is estimated by varying the
mistag fractions in each $r$ interval
within their uncertainties.
A possible fit bias is estimated from a large sample of MC simulated signal decays.
The effect of the peaking background is studied using MC simulations
allowing for $CP$ violation in non-resonant decays.
The possible interference between Cabibbo-favored $b \to c \bar{u} d$
and suppressed $\bar{b} \to \bar{u} c \bar{d}$ amplitudes
in the decay of the tagging $B$ meson,
referred to as tag-side interference~\cite{Long2003_tagside_interference}, is studied using MC simulations
with inputs obtained from $B^{0} \to D^{*-} \ell^{+} \nu_{\ell}$ control samples.
The largest deviations in the above MC studies are assigned as systematic uncertainties.
The total systematic uncertainty is obtained by adding all contributions in quadrature.

The significance of the results is studied by a likelihood-ratio approach.
For $B^{0} \to D^{+}D^{-}$ we exclude the conservation of $CP$ symmetry ($\mathcal{S}_{D^{+}D^{-}} = \mathcal{C}_{D^{+}D^{-}} = 0$)
at a confidence level of $1 - 2.7 \times 10^{-5}$ corresponding to $4.2\sigma$.
For $B^{0} \to D^{*\pm}D^{\mp}$ the conservation of $CP$ symmetry ($\mathcal{A}_{D^{*}D} = \mathcal{S}_{D^{*}D} = \mathcal{C}_{D^{*}D} = 0$)
is excluded at a confidence level of $1 - 6.8 \times 10^{-5}$ corresponding to $4.0 \sigma$.
These results account for both the statistical and the systematic uncertainties.

The fit procedure was validated by various cross-checks.
The same analysis was performed for $B^{0} \to D_{s}^{+} D^{(*)-}$ decays.
The results are 
$\mathcal{A}_{D_{s} D} = -0.01 \pm 0.02$, $\mathcal{S}_{D_{s} D} = -0.05 \pm 0.05$, $\mathcal{C}_{D_{s} D} = +0.01 \pm 0.03$, $\Delta\mathcal{S}_{D_{s} D} = +0.01 \pm 0.05$ and $\Delta\mathcal{C}_{D_{s} D} = -0.95 \pm 0.03$
in $B^{0} \to D_{s}^{+} D^{-}$ 
and
$\mathcal{A}_{D_{s} D^{*}} = +0.01 \pm 0.02$, $\mathcal{S}_{D_{s} D^{*}} = -0.04 \pm 0.05$, $\mathcal{C}_{D_{s} D^{*}} = +0.06 \pm 0.03$, $\Delta\mathcal{S}_{D_{s} D^{*}} = +0.10 \pm 0.05$ and $\Delta\mathcal{C}_{D_{s} D^{*}} = -1.00 \pm 0.03$
in $B^{0} \to D_{s}^{+} D^{*-}$,
where the uncertainties are statistical only.
The results are consistent with the assumption of no $CP$ violation in $B^{0} \to D_{s}^{+} D^{(*)-}$ decays.
The lifetimes determined by fits to untagged $B^{0} \to D^{+}D^{-}$ and $B^{0} \to D^{*\pm}D^{\mp}$ samples
are consistent with the world average~\cite{PDG}.

In summary we report measurements of the branching fractions and time-dependent $CP$ violating asymmetries in
$B^{0} \to D^{+}D^{-}$ and $B^{0} \to D^{*\pm}D^{\mp}$ decays using the final Belle data sample of $( 772 \pm 11 )\times 10^6 B\bar{B}$ pairs.
We measure the branching fractions
$\mathcal{B}\left(B^{0} \to D^{+}D^{-}\right)= \left( 2.12 \pm 0.16 \pm 0.18 \right)\times 10^{-4}$
and
$\mathcal{B}\left(B^{0} \to D^{*\pm}D^{\mp}\right)= \left( 6.14 \pm 0.29 \pm 0.50 \right)\times 10^{-4}$.
The measured $CP$ asymmetry parameters are
$\mathcal{S}_{D^{+}D^{-}} = -1.06_{\ -0.14}^{\ +0.21} \pm 0.08$ and
$\mathcal{C}_{D^{+}D^{-}} = -0.43 \pm 0.16 \pm 0.05$
in $B^{0} \to D^{+}D^{-}$
and
$\mathcal{A}_{D^{*}D}       =  +0.06 \pm 0.05 \pm  0.02$,
$\mathcal{S}_{D^{*}D}       =  -0.78 \pm 0.15 \pm  0.05$,
$\mathcal{C}_{D^{*}D}       =  -0.01 \pm 0.11 \pm  0.04$,
$\Delta\mathcal{S}_{D^{*}D} =  -0.13 \pm 0.15 \pm  0.04$ and
$\Delta\mathcal{C}_{D^{*}D} =  +0.12 \pm 0.11 \pm  0.03$
in $B^{0} \to D^{*\pm}D^{\mp}$. 
For $B^{0} \to D^{+}D^{-}$, the $CP$ asymmetries are approximately $0.5\sigma$ outside
of the physical parameter space
defined by $\sqrt{ \mathcal{S}_{D^{+}D^{-}}^{2} + \mathcal{C}_{D^{+}D^{-}}^{2} } \leq 1$
and the direct $CP$ asymmetry deviates from zero by approximately $2.0\sigma$.
For $B^{0} \to D^{*\pm}D^{\mp}$, if the contribution of penguin diagrams is negligible
and if the hadronic phase between $B^{0} \to D^{*+} D^{-}$ and $B^{0} \to D^{*-} D^{+}$ amplitudes is zero
and their magnitudes are the same,
then $\mathcal{A}_{D^{*}D}$, $\mathcal{C}_{D^{*}D}$, $\Delta\mathcal{S}_{D^{*}D}$ and $\Delta\mathcal{C}_{D^{*}D}$ vanish
and $\mathcal{S}_{D^{*}D}$ is equal to $\sin{2\phi_1}$.
Our result is consistent with the above and we measure $\sin{2\phi_1} = -0.78 \pm 0.15 \pm  0.05$.
The $CP$ asymmetries obtained in $B^{0} \to D^{+}D^{-}$ and $B^{0} \to D^{*\pm}D^{\mp}$ decays are
both in agreement with measurements of decays involving $b\to(c\bar{c})s$ transitions~\cite{BaBar_btoccs,Belle_btoccs}
and with previous measurements of $B^{0} \to D^{(*)\pm}D^{(*)\mp}$ decays~\cite{Aushev2004, Fratina2007, Vervink2009, Aubert2009_btodd}.
We find evidence for $CP$ violation in both decay channels with a significance of $\geq$4$\sigma$.
These results supersede previous measurements of branching fractions and time-dependent $CP$ asymmetries in
$B^{0} \to D^{+}D^{-}$ and $B^{0} \to D^{*\pm}D^{\mp}$ by the Belle Collaboration~\cite{Fratina2007,Aushev2004,BF_DstarD_Belle}.

We thank the KEKB group for excellent operation of the
accelerator; the KEK cryogenics group for efficient solenoid
operations; and the KEK computer group, the NII, and 
PNNL/EMSL for valuable computing and SINET4 network support.  
We acknowledge support from MEXT, JSPS and Nagoya's TLPRC (Japan);
ARC and DIISR (Australia); NSFC (China); MSMT (Czechia);
DST (India); INFN (Italy); MEST, NRF, GSDC of KISTI, and WCU (Korea); 
MNiSW (Poland); MES and RFAAE (Russia); ARRS (Slovenia); 
SNSF (Switzerland); NSC and MOE (Taiwan); and DOE and NSF (USA).


\begin{thebibliography}{99}
\bibitem{phione} Another naming convention, 
$\beta$ ($= \phi_1$), is
also used in the literature.
\bibitem{CC}
In this article the inclusion of charge conjugated decay modes
is implied unless otherwise stated.
\end{thebibliography}
\end{document}